# The Impact of LIBOR Linked Borrowing to Cover Venture Bank Investment Loans Creates a New Systemic Risk.

Brian P. Hanley

Brian.Hanley@ieee.org

## Abstract

A scenario in which regulators take the drastic step of requiring coverage of all venture-bank investment loans using interbank borrowed funds is considered. In this scenario, to attempt to compensate for the cost, a minimal amount of equity default clawback swap coverage is used to ensure that Tier 1 and 2 capital requirements are still met. To do this, the equity default swap percentage on all investment loans is cut to between 3.88% and 2.88%.

Results: For a portfolio of 1.31X (ten year total conventional return) or better, at interest rates of 2% or less, the venture-bank can survive and have good returns. For a portfolio of 1.5X (ten year total conventional return) the bank can have extraordinary returns if the interest paid is below 1.5% and survive up to 3%. interest. However, if returns fall from this level (which is among the best such large portfolios), or interest rates rise, then venture-banks fail. Since the lower half of VC funds are at or below 1X 10 year return, and LIBOR rates have spent long periods of most decades above 2%, venture-banking as a whole would not survive. In the median LIBOR rate scenario of the period from 1994 to 2014, no large portfolio entity would survive. All would fail.

Conclusion: Requiring use of LIBOR funds limits profitability and damages stability of the venture-banks, with no visible benefit to any party, while creating a new systemic risk to the larger banking system that should have very serious repercussions.

## 1 Introduction

The Equity Default Clawback Swap (EDCS) based banking venture-banking system has already been laid out (Hanley, 2017a, 2017b). To fully understand this discussion, it is necessary to understand the venture-bank system as proposed in those papers, with the core of it contained in the former.

In brief, this system is based on the fundamental algorithm where:

***A.** A bank issues a loan to a borrower.*

***B.** The bank purchases a derivative to act as insurance against default on the loan.*

***C.** The bank then books this insured valuation of the loan into its capital account.*

This allows the bank to issue a new loan up to the amount of the insurance policy without needing to go to the central bank to acquire more reserves (Hanley, 2012). Herein I will discuss the impact of what I would consider a regulatory failure, or perhaps better termed, sabotage of the venture-banking system.

I do not think that a properly run venture-banking system as I define it will increase risk, I think it will lower it because it creates a new money-creation system decoupled from the larger banking system except for settlements, which modeling shows to be very profitable. The profits go into the larger system when exits occur, while the losses are contained within the venture-bank system. Therefore, I do not believe that regulators should change regulations to force coverage of insured loan assets with borrowed interbank funds. However, not all regulatory moves are done for the benefit of the system. It is conceivable that some entity could lobby to essentially force double-coverage of the EDCS backed loan assets in order to make a short term profit from the losses that would result from breaking the system.

## 2 Impact of double-coverage of insured assets with short-term funds

### 2.1 Overview

What would happen in this scenario is that regulators would require a venture-bank to cover its insured asset loans with deposits or borrowed funds. To do that, the venture-bank would need to borrow on the open market. To model the impact of this, cost is based on records of the London interbank offered rate (LIBOR) (FRED 2016). Note here that these costs could easily be higher than LIBOR.

The modeling results shown below are also made assuming long-term (e.g. lifetime of the venture-bank) in order to be conservative. However, as we will see, such a regulatory move would be a short-term event that may last 6-24 months before the venture-banks crashed. Again, I do not think this is a valid regulatory decision. However, the question came up, so I modeled it to see if the venture-bank system could survive with such a burden. It is possible, but only for a small fraction of venture banks.

### 2.2 Interest rate data source – LIBOR

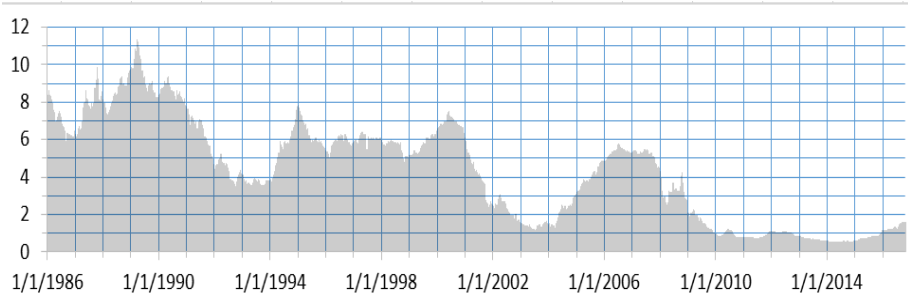

*Figure 1: LIBOR – (1986-2016) 30 year median 4.26%, mean 4.58%. (1996-2016) 20 year median 2.44%, mean 3.10%. (2006-2016) 10 year median 1.06%, mean 1.94%*

Looking back to 1986, the median and mean LIBOR were 4.26% and 4.58% respectively, which would yield funds borrowing rates of 4.51% and 4.83% in this model. These data are used in modeling of the LIBOR dependent version.

### 2.3 Net return data source – Kauffman's set of 99 venture capital firms

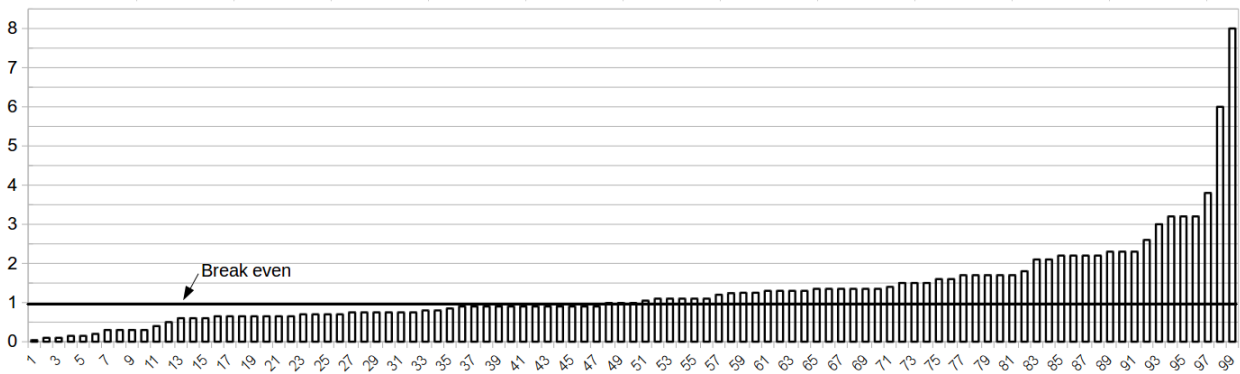

*Figure 2: Kauffman venture capital fund dataset. (Mulcahy, 2012)*

Each bar in figure 2 is return for one of the 99 venture capital firms, net of VC firm's 2% fee and 20% carry. My model uses a compressed version of this with the same overall characteristics obtained by averaging pairs. By adding and subtracting from this compressed dataset, varying rates of return are modeled, all with the same overall spread as Kauffman's dataset.

### 2.4 EDCS coverage for risk mitigation of interbank lenders

EDCS coverage is the percentage of each investment in a loan portfolio that is covered by a EDCS. This would be determined by two primary factors as mentioned in section 2 above: First, that a venture-bank must maintain a sufficient level of insurance on its portfolio that pulling the plug on part of its portfolio will not leave it scrambling to meet Tier 1 and Tier 2 reserve requirements.  Second, that a venture-bank must also maintain a sufficient level of insurance to ensure that interbank lenders are able to provide preferred loan rates without undue systemic risk. It is this latter requirement that is likely to set the minimum level of EDCS coverage at between the 2.88% floor, and 3.88%.

**2.5 Calculation of EDCS coverage levels to mitigate risk of interbank lenders**

I used the historical Kauffman data shown in figure 2, and approached the problem in two ways. For the first method, I used a cutoff based on standard deviation. The Kauffman data has a standard deviation of 1.116 over 99 venture funds. To generate a conservative number, I reset all data points more than one standard deviation above break-even to the break-even value of 1 (one). This generated a net loss on the total portfolio of 2.72%. This projects a minimum EDCS coverage rate of 5.6%, obtained by summing the 2.88% required for reserves maintenance, and 2.72% needed for interbank lenders. Using this value shifts venture-bank profitability down somewhat. At a 2% inter-bank funds rate, the 1.31X portfolio would become 1.50 at a 30X MOC, and 2.15 at the 43X MOC down approximately 0.45X.

For the second method, I set all of the portfolio returns above the break-even value of 1 (one) to breakeven (e.g. 1). This generated a highly conservative loss projection. Using this method, the net portfolio loss was 17.45%. Summing 2.88% and 17.45% generates a ceiling of 20.33% EDCS coverage. This level of EDCS coverage will only work with a rate of return of 1.47X or higher at the 30X MOC level. I do not think this would allow the venture-bank business as a whole to survive.

The actual decisions on EDCS coverage would be up to the interbank lenders. However, it should be possible to see how I chose the value of 3.88% EDCS coverage as reasonable. Certainly for established parties that can show history, given the interesting conclusion that past performance had significance predicting future results in venture capital (Mulcahy, 2012), 3.88% should provide adequate coverage for both requirements.

**2.6 Portfolio models with 1.82% LIBOR funds rate, 3.88% EDCS coverage and 5% EDCS premiums**

Model results below are using a 1 year bank rate of 1.82%, for three different portfolios, high (1.50 return), medium (1.31 return) and low (1.10 return). However, note that the median portfolio per Mulcahy, (figure 1) has a return of 1.) Current 12 month LIBOR is 1.57%, so the bank rate is set at LIBOR + 0.25%. These three portfolios have 1.50, 1.31 and 1.10 net returns when operated conventionally.

EDCS coverage is set at 3.88% of each investment. This level is 1.347 times the minimum that is required to meet tier capitalization requirements, and is chosen as a conservative value to ensure a venture-bank doesn't have a reserves crisis in case of failure of investments. It is also chosen as a level that I believe should be sufficient for most venture-banks in order to secure preferred interbank loan rates.

In this 10 year to exit model, the EDCS rate is fixed at 5%, and pay-off year is set at year 5. After paying off on a EDCS, there is no more premium income and the bank rate is used to determine the cost of money to carry the pay-off amounts for the next 5 years.

Note that in these models I keep a conservative margin on the maximum capital, calling out the 30X and 43X multiples of capital. Portfolio sensitivity to LIBOR with 3.88% of invested funds covered by EDCS

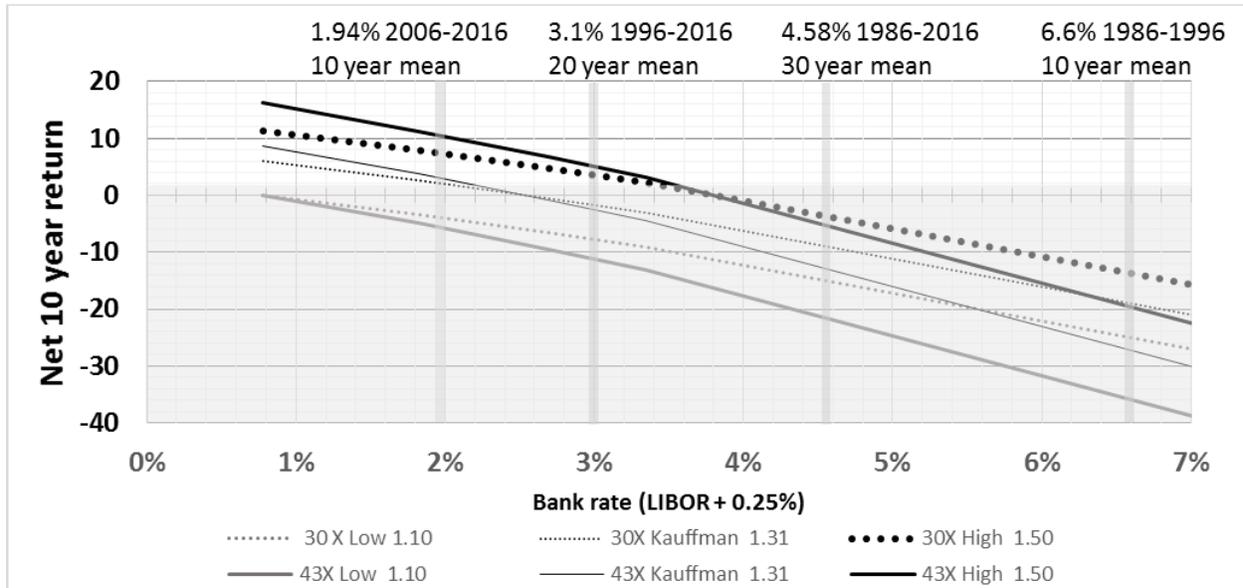

*Figure 3: Venture bank sensitivity to interest rate by classical rate of return. Break-even at 1.0. These data points are MOC's of 30X and 43X for each of the three portfolios given historical LIBOR values for the period from 1996 to 2016 (FRED, 2016). During this period, the LIBOR high was 7.50%, the low was 0.53%. Modeling data for venture capital returns from Mulcahy. (Mulcahy, 2012)*

*2.6.1 EDCS return sensitivity to LIBOR with 3.88% of invested funds covered by EDCS*

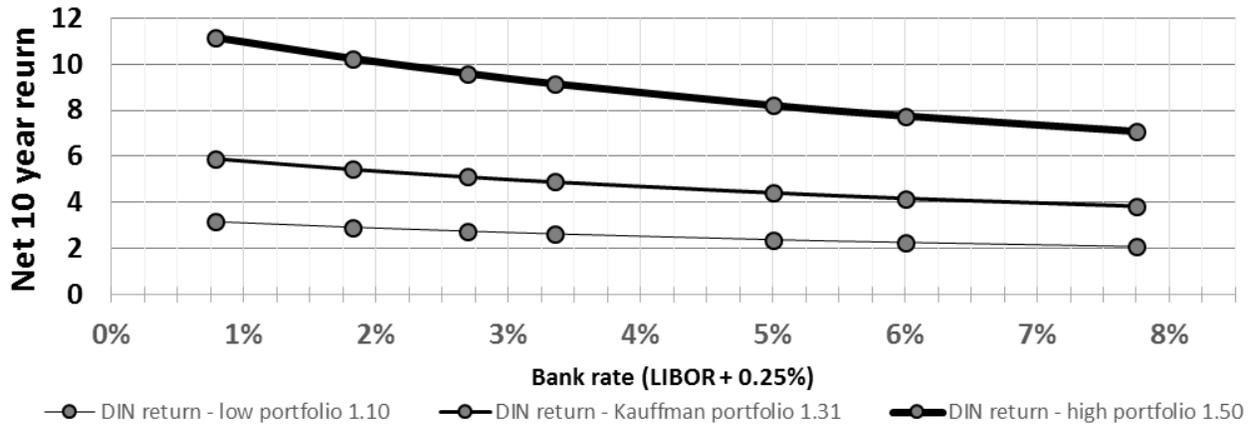

*Figure 4: EDCS business total gross rate of return variability by interest rate. Breakeven = 0*

## 3 Concluding remarks

The impact of a long-term requirement to replace EDCS insured loans with LIBOR based coverage would be severe on the venture-banks. Except when LIBOR rates are at historical minima, in the long-term most venture-banks would be underwater if their conventional VC returns were less than 1.3 or so. (Fig. 3)

Underwriters should make money, as long as their venture-bank clients didn't go under. (Fig. 4)

Should regulators take this stance, it should be understood that it would be, essentially, deliberately destructive. It could not persist over the long-term within that jurisdiction. To survive, the only response could be to transfer all assets to another jurisdiction that did not require it.

This regulatory decision would dramatically lower the stability and profitability of venture-banks. Worse, it creates a link into the larger banking system consisting of borrowed LIBOR funds that would disappear when the venture-banks failed. This creates systemic risk where none would exist when using default insurance alone, because default insurance backed capital assets are decoupled from the rest of the banking system.

## 4 Glossary

EDCS – Equity Default Clawback Swap. A proposed derivative that insures loans made by venture capitalists as investments.

FRED – Federal Reserve Economic Data.

LIBOR – London Interbank Offered Rate. For this use, it means the 12 month rate.

MOC – Multiple of Original Capital. The total outstanding investments divided by the original capital placed in bank Tier 1 reserves is the MOC.

## 5 References

Bank for International Settlements (2016) Basel Committee on Banking Supervision. Basel, Switzerland.

Division of Banking Supervision and Regulation (2016) Commercial Bank Examination Manual. Washington, DC: Board of Governors of the Federal Reserve System.

FRED (2016) 12-Month London Interbank Offered Rate (LIBOR), based on U.S. Dollar. Federal Reserve Economic Data, Federal Reserve of St. Louis. https://fred.stlouisfed.org/series/USD12MD156N

Hanley, B.P. (2012). Release of the Kraken: A Novel Money Multiplier Equation's Debut in 21st Century Banking. Economics, 6:3.

Hanley, B.P. (2017a). Equity Default Clawback Swaps to Implement Venture Banking. arXiv:1707.08078 [q-fin.GN].

Hanley, B.P. (2017b). The perverse incentive for default swap instruments that are derivatives: solving the jackpot problem with a clawback lien for equity default swaps. arXiv:1711.02600 [q-fin.GN]

Mulcahy, D., Weeks, B., Bradley, H.S. (2012) 'We have met the enemy…and he is us' Lessons from Twenty Years of the Kauffman Foundation's Investments in Venture Capital Funds and The Triumph of Hope over Experience. Ewing Marion Kauffman Foundation.